\documentclass[lettersize,journal]{IEEEtran}
\usepackage{amsmath,amsfonts}
\usepackage{algorithmic}
\usepackage{algorithm}
\usepackage{array}
\usepackage[caption=false,font=normalsize,labelfont=sf,textfont=sf]{subfig}
\usepackage{textcomp}
\usepackage{stfloats}
\usepackage{url}
\usepackage{verbatim}
\usepackage{graphicx}
\usepackage{cite}
\begin{document}

\title{Computational Microwave Localization and Material Identification Using Conformal Metasurface Antennas}
\author{Sajedeh~Keshmiri, \IEEEmembership{Graduate Student Member, IEEE}, and Mohammadreza F. Imani \IEEEmembership{Member, IEEE}
\thanks{The authors are with the School of Electrical, Computer, and Energy Engineering, Arizona State University, Tempe, AZ 85287 USA (e-mail: s.keshmiri@asu.edu; mohammadreza.imani@asu.edu).}
\thanks{}}

\markboth{Journal of \LaTeX\ Class Files,}%
{Shell \MakeLowercase{\textit{et al.}}: A Sample Article Using IEEEtran.cls for IEEE Journals}


\maketitle

\begin{abstract}
This paper presents a simple and compact microwave sensing approach using conformal frequency-diverse metasurfaces to jointly localize and identify materials within a pipe. The frequency-diverse metasurfaces consist of an electrically large SIW patterned with metamaterial radiators with diverse resonant frequencies. These antennas can generate spatially distinct patterns that can encode information into simple frequency sweeps. Two such antennas are wrapped around a pipe to localize and identify objects inside, thereby replacing large, complex tomographic antenna arrays with simple frequency sweeps. To do that, a sensing matrix is experimentally populated using targets made of different materials placed in
a grid of possible locations. Using computational processing, we show that the system successfully distinguishes among metal, wood, and nylon rods and accurately localizes their positions using unique frequency signatures. The extension of this method to two-object localization is examined. The results highlight a simple, low-cost, and reliable framework for microwave sensing that enables material characterization and localization without large arrays, scanning, or complex hardware.
\end{abstract}


\section{Introduction}
\IEEEPARstart{M}{icrowave} sensing has become an indispensable technology for non-destructive evaluation (NDE), material characterization, and object localization due to its ability to penetrate optically opaque media and capture both electromagnetic and geometric properties of targets\cite{zoughi2000microwave,pastorino2010microwave}. Traditional microwave imaging and sensing systems often rely on large antenna arrays, mechanical scanning, or complex switching networks, which increase hardware complexity, cost, and data acquisition time \cite{sheen2002three,moulder2016development,ahmed2012advanced}. These limitations motivate the development of compact sensing platforms capable of extracting spatial information using a small number of measurements.

The complexity of traditional hardware stems from the fact that it yields a one-to-one mapping between measurement and the target under test. Recent advances in compressive sensing have shown that spatial information can be encoded in a few measurements and recovered through signal processing. This idea has led to methodologies in which the antenna hardware and processing techniques are co-designed to reduce hardware complexity by shifting the burden of sensing from hardware to software. Using this method, novel planar microwave imaging configurations for a variety of applications, such as security screening\cite{gollub2017large,imani2020review}, gesture recognition\cite{li2019machine}, or through-the-wall imaging \cite{sleasman2019computational}, have been demonstrated. It is shown that by utilizing metasurface antennas that can generate spatially diverse patterns as a function of frequency or electronic signals, the information from the entire scene can be multiplexed into a few measurements collected by a few antennas. While the metasurface antennas no longer yielded a one-to-one mapping, computational processing their collected data enabled the recovery of high-quality images of large objects (e.g., the human body) at video frame rates \cite{gollub2017large,imani2020review}.

\begin{figure*}[!t]
\centering
\includegraphics[width=0.8\linewidth]{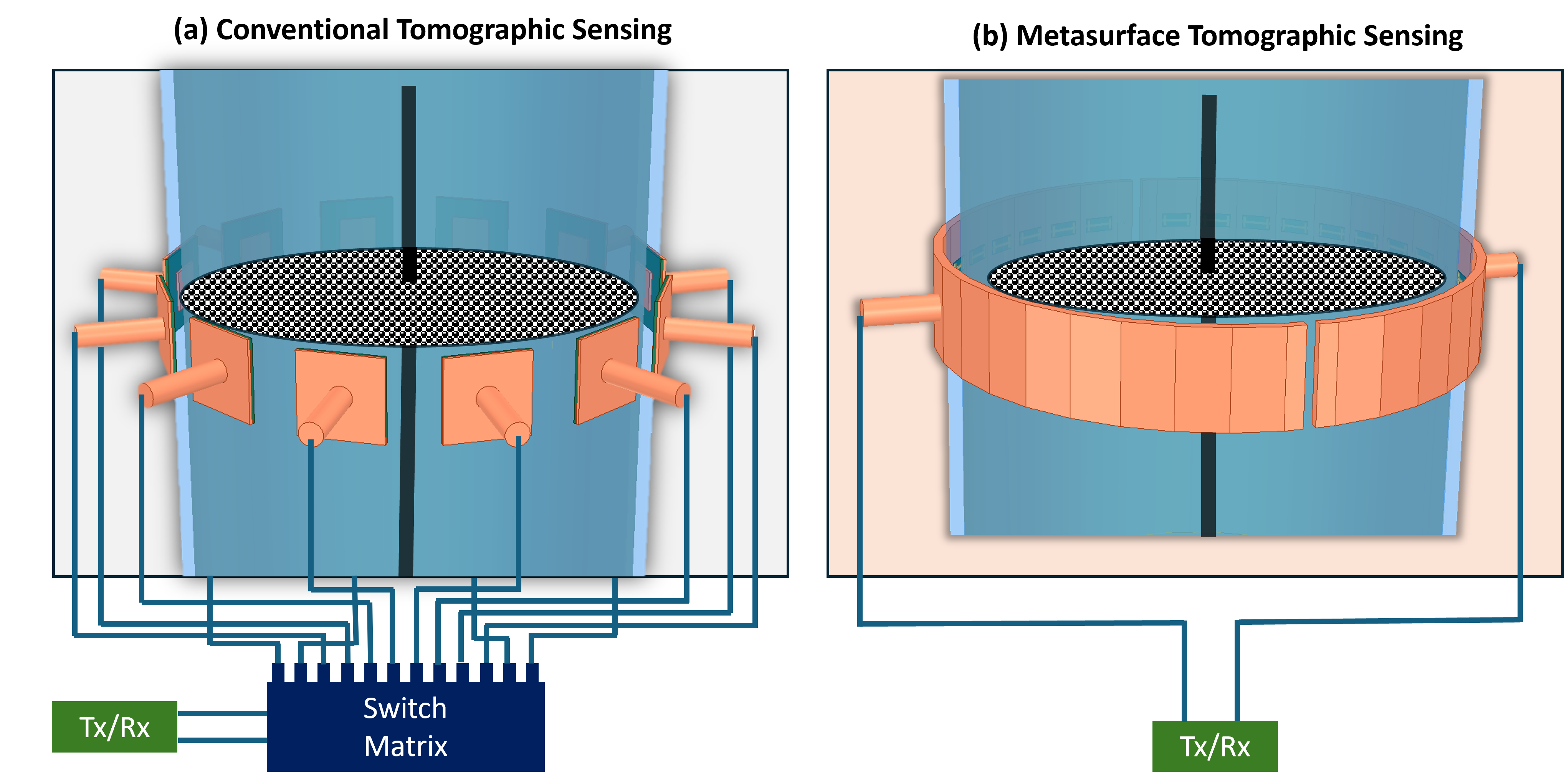}
\caption{(a) Conventional tomographic sensing using antenna arrays and a switch matrix. (b) Proposed sensing method using two conformal frequency-diverse antennas.}
\label{fig:fig2}
\end{figure*}

Extending this idea to tomographic configurations has seen less exposure. In such a configuration, a large antenna array connected to a costly switch matrix is used to detect anomalies or objects within trees, concrete columns, pipes, grain silos, or human breasts, necks, and heads \cite{li2003conformal,kim2003microwave,boero2018microwave,doroshewitz2019microwave,amineh2024microwave,carrigan2018nondestructive,stakenborghs2009microwave,wu2020microwave,amineh2020nondestructive,lovetri2020innovations,dachena2022initial}. The large number of antennas, however, makes these methods complicated and time consuming. To overcome this challenge, significant advances have been made in reducing the number of antennas or measurements required \cite{oliveri2019compressive,ambrosanio2014compressive,park2019real}. However, the main hardware still consists of a large antenna array (albeit with fewer antennas or fewer antenna-pair measurements) connected to a switch matrix.

We approach the problem of tomographic microwave sensing from a different perspective, in which no antenna array or switch matrix is required, as illustrated in Fig.~\ref{fig:fig2}. Instead, we plan to utilize two multiplexing metasurface antennas (one as a transmitter and one as a receiver) that can conform to the object surface. They generate spatially diverse patterns that illuminate the whole object and multiplex its information into simple transmission measurements. More specifically, we use a large metasurface antenna patterned with metamaterial elements whose resonant frequencies are randomly selected from a band of operation. This antenna can generate distinct radiation patterns as a function of frequency. When used in the proposed conformal setup, these multiplexing antennas effectively replace the large antenna arrays with frequency sweeps. In such approaches, distinct frequency-dependent signatures are associated with different object locations and/or materials, enabling localization through cross-correlation examination. This is to some extent similar to the idea of using unique frequency responses to detect objects inside a room \cite{wu2015non,kotaru2015spotfi,ma2019wifi,delhougne2018precise,del2018dynamic}. Instead of leveraging the room's multiple scattering to obtain unique signatures, we use conformal frequency-diverse antennas. 

\begin{figure*}[!t]
\centering
\includegraphics[width=0.6\linewidth]{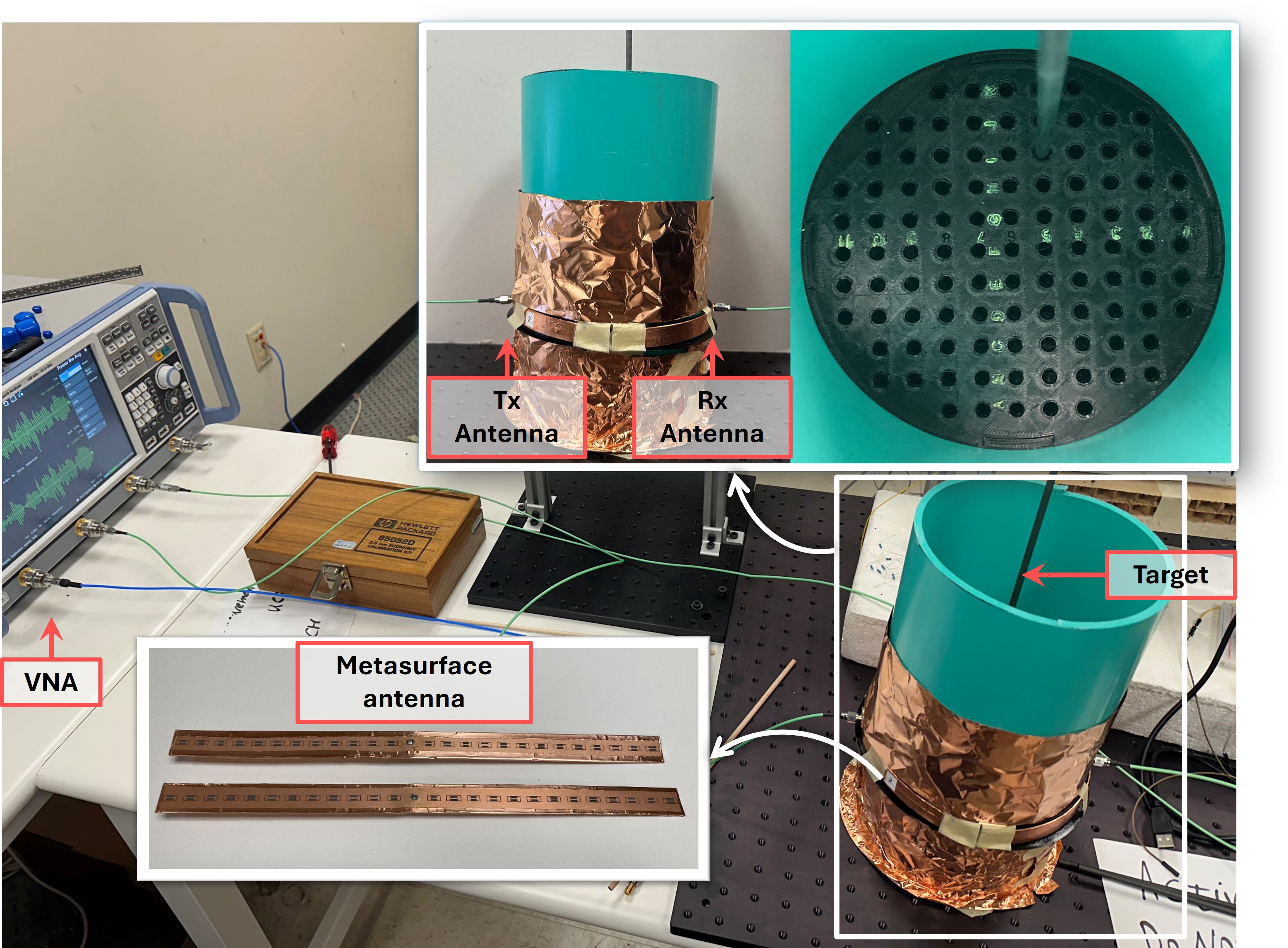}
\caption{Experimental measurement setup including the VNA, metasurface antennas mounted on the pipe structure, and the target placement region. Copper sheets were added to reduce interference from nearby measurements and human movement. They are not essential for sensing and are included only in lab-setting measurements.}
\label{fig:fig3}
\end{figure*}

In this paper, we demonstrate the proposed microwave sensing framework and show its ability to simultaneously classify materials and localize them. Building on our preliminary results reported in \cite{11317837}, we first construct a sensing matrix that encodes candidate locations and materials of interest. Using this matrix, we develop a computational method to detect both the location and the material. The impact of frequency bandwidth on the detection capabilities is examined. We show that this method can possibly be extended to more than one object. This framework provides a simple and scalable solution for microwave sensing applications such as non-destructive evaluation, material characterization, and smart sensing environments without requiring large arrays or mechanical scanning.

The contributions of this paper, beyond those in \cite{11317837} and other previous works, are as follows: 1) A compact computational microwave sensing framework based on conformal frequency-diverse metasurface antennas is experimentally demonstrated as a novel alternative to large tomographic structures for simultaneous material identification and spatial localization. 2) A sensing methodology based on the cross-correlation of a sensing matrix consisting of transmission measurements is developed to distinguish both target material and position using simple frequency sweeps.
3) The impact of frequency diversity and bandwidth on sensing performance is experimentally investigated for different numbers of frequency points.
4) A leave-one-out measurement strategy is conducted to validate the robustness and generalization capability of the proposed sensing framework without data reuse.
5) Proof-of-concept two-object detection is demonstrated using a CGS-based reconstruction approach, showing the potential extension of the proposed framework to more complex sensing scenarios.

\section{Proposed computational sensing system}
\subsection{Metasurface Antenna}
\noindent The sensing platform is implemented using a pair of conformal, frequency-diverse metasurface antennas operating in the X-band. The design of these antennas has been detailed in \cite{alamzadeh2025experimental} and is not repeated here. Instead, we review only its key features. These antennas are fabricated on 30-mil Rogers/Duroid 5880, which can conform to curved geometries. In this work, we bend them to circumvent a cylindrical pipe. Each antenna consists of a substrate-integrated waveguide (SIW) (width of 15~mm, a guided-mode cutoff frequency of 6.74~GHz, and a length of 141.4~mm) loaded with 17 metamaterial radiating elements distributed along its surface. The geometry of these elements is intentionally varied so that each resonates at a slightly different frequency within the X band. Consequently, the antenna produces spatially diverse radiation patterns that change with frequency. In \cite{alamzadeh2025experimental}, the conformal metasurfaces were used for computational angle-of-arrival detection\cite{imani2023conformal}. Here they are repurposed for tomographic sensing within a cylindrical pipe. Instead of detecting external incident waves, the system leverages its frequency-diverse near-field distribution to probe and characterize objects inside the pipe.

\begin{figure}[t]
\centering
\includegraphics[width=0.8\columnwidth]{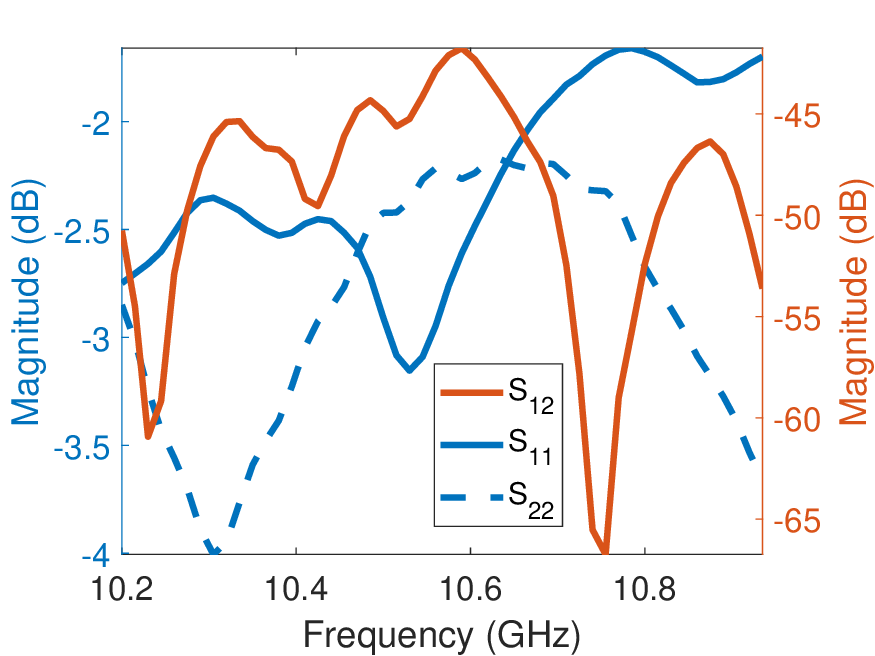}
\caption{Measured $S$-parameters of the antenna–pipe structure with no object present in the sensing region.}
\label{fig:fig4}
\end{figure}

For this study, two similar metasurface antennas are mounted on opposite sides of a PVC pipe with an outer diameter of 90 mm. The antennas are wrapped around the exterior surface and connected via coaxial SMA ports to VNA ports 1 and 2. Copper tape is used to implement the SIW sidewalls in this proof-of-concept demonstration. The overall experimental setup, including the VNA, conformal metasurface antennas, and target placement region, is shown in Fig.~\ref{fig:fig3}. In Fig.~\ref{fig:fig4}, we have reported the measured S-parameters of the antennas when mounted around the (empty) pipe structure. They clearly exhibit frequency-dependent variations over the operating band: as the excitation frequency is swept, the effective radiation distribution along the metasurface changes, producing distinct electromagnetic field patterns within the sensing region. Objects placed within this region perturb the electromagnetic field distribution in a manner that depends on both their material properties (permittivity and conductivity) and their spatial location. These perturbations are directly reflected in the measured transmission response.

\subsection{Sensing Procedure}
\noindent To sense objects placed inside the pipe, we utilize the transmission coefficient (i.e. $S_{12}$) between the two metasurfaces. To ensure controlled and repeatable target placement, a 3D-printed positioning grid is inserted inside the pipe. The grid contains 101 predefined holes arranged in a uniform lattice, allowing cylindrical rods to be placed at known spatial locations. The diameter of each hole (or target) is 5.5~mm, and the center-to-center spacing between adjacent holes is 12~mm ($\approx 0.18\lambda$ and $\approx 0.4\lambda$, respectively, at 10~GHz). Unlike the previous study in \cite{11317837}, which considered only a metallic rod, the present work includes rods made of metal, wood, and nylon with similar geometric dimensions. This allows the system to capture variations in electromagnetic response due to both spatial position and material properties. 

First, a background response is recorded with no object inside the pipe. This reference measurement captures the coupling between the two metasurface antennas. An object is then inserted at a specific grid location, and the transmission response is recorded at 401 uniformly spaced points over 9 to 12 GHz. We start the measurement process with a large bandwidth and a large number of frequency points. Depending on the sensing task at hand, we may only need a subset of all measurements (or bandwidth). The background measurement is subtracted from the object-present measurement to isolate the perturbation introduced by the target. To account for potential variation in the exact measurement spot in practice, each configuration (position--material pair) is measured 5 times. For each measurement, we removed the rod and returned it to the same hole. As a result, the five measurements differed slightly, accounting for variation that can happen in practice.

 Next, we define a sensing matrix $\mathbf{H}$ whose columns correspond to the frequency-domain signatures associated with each position-material combination. The column of $\mathbf{H}$ corresponding to the $j$ location is populated by averaging the 5 measurements discussed above and subtracting the background response. This averaging reduces measurement noise and improves the robustness of the sensing dictionary to possible shifts from the grid. Mathematically, this process is described as:
\begin{equation}
H_{ij} = \overline{S_{12}^{(j)}(f_i)} - S_{12}^{bg}(f_i),
\end{equation}
where $\overline{S_{12}^{(j)}(f_i)}$ represents the average transmission response obtained from repeated measurements for the $j^{\text{th}}$ configuration, and $S_{12}^{bg}(f_i)$ denotes the background response at frequency $f_i$ when no object is present. Similarly, the background-subtracted measurement vector is defined as
\begin{equation}
g_i = S_{12}(f_i) - S_{12}^{bg}(f_i),
\end{equation}

The detection is performed using a correlation-based approach, where the measured vector $\mathbf{g}$ and the columns of the sensing matrix $\mathbf{H}$ are first centered and normalized. Centering is performed by subtracting the mean value from $\mathbf{g}$ and from each column of $\mathbf{H}$, which removes any constant offset in the measurements. The resulting vectors are then normalized by their $\ell_2$ norms to ensure that the correlation depends only on the similarity of their signatures, not on their magnitudes. The correlation score is then computed as $|\mathbf{H}_n^H \mathbf{g}_n|$ where $(\cdot)^H$ denotes the Hermitian (conjugate transpose) operation. The index corresponding to the maximum value determines the estimated target location and material.

The sensing process can also be expressed within a linear model framework. Given a measurement vector $\mathbf{g}$, the detection problem can be expressed as
\begin{equation}
\mathbf{g} = \mathbf{H}\mathbf{F},
\end{equation}
where $\mathbf{F}$ represents the spatial--material distribution vector. For single-object detection, $\mathbf{F}$ contains a single nonzero entry corresponding to the correct position and material class.
For multi-object scenarios, where the measurement is a superposition of multiple dictionary elements, the inverse problem is solved using the conjugate gradient squared (CGS) algorithm, as discussed in Section III-B.

\begin{figure}[t]
    \centering

    \begin{minipage}[t]{0.8\columnwidth}
        \centering
        \rlap{\raisebox{-5pt}{\scriptsize\textbf{(a)}}}
        \includegraphics[width=\linewidth]{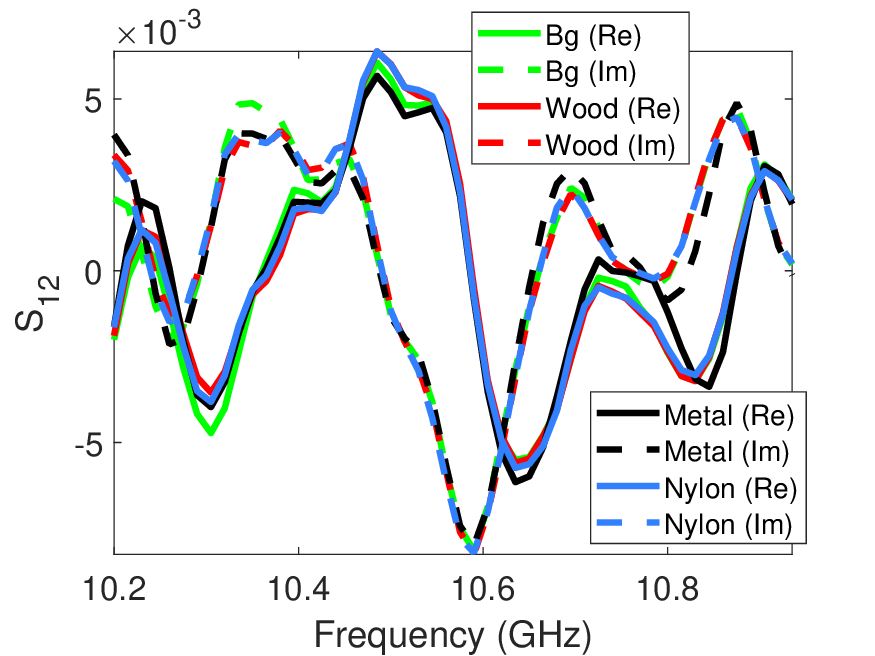}
    \end{minipage}
    \hfill
    \begin{minipage}[t]{0.8\columnwidth}
        \centering
        \rlap{\raisebox{-5pt}{\scriptsize\textbf{(b)}}}
        \includegraphics[width=\linewidth]{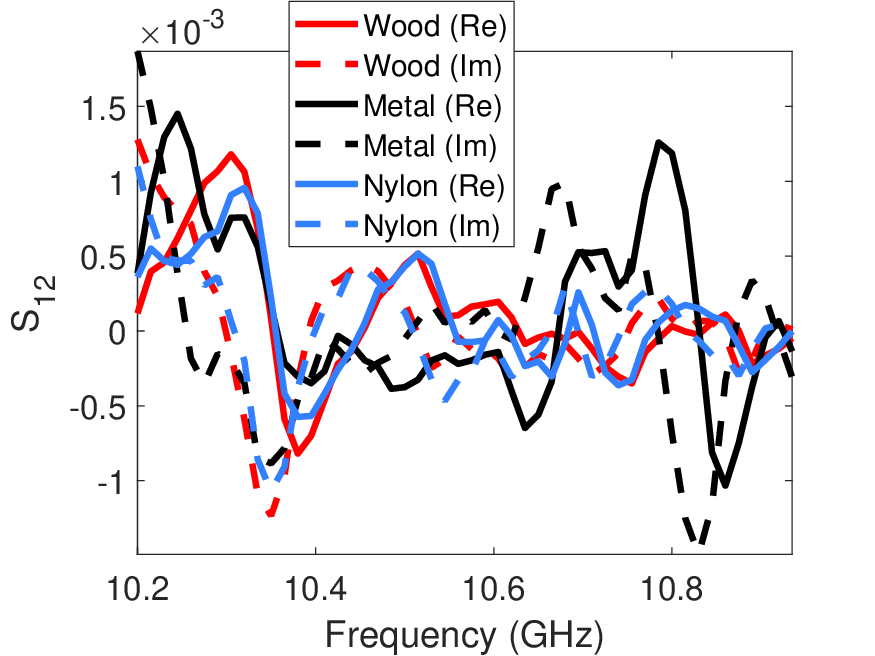}
    \end{minipage}

    \caption{(a) Measured $S_{12}$ for the background (no object) and for metal, wood, and nylon targets placed at the center of the sensing region. (b) Differential $S_{12}$ obtained by subtracting the background response from the corresponding object measurements.}
    \label{fig:fig5}
\end{figure}

\section{Results}
\subsection{Material and Position Detection}

Fig.~\ref{fig:fig5}(a) illustrates the measured $S_{12}$ response for the background (no object) and for metal, wood, and nylon rods placed at the center of the pipe.  It is worth noting that, for better visualization, we plotted these measurements only over a subset of the full bandwidth. We observe small perturbations, which are expected given the subwavelength size of the target (diameter of $\approx 0.18\lambda$ at 10~GHz). To more clearly isolate the perturbation introduced by the objects, we subtract the background response from the measurements. The resulting differential responses are shown in Fig.~\ref{fig:fig5}(b). After subtraction, the signatures associated with each material become more pronounced, revealing distinct frequency-dependent patterns for metal, wood, and nylon. These unique signatures form the basis for constructing the sensing matrix and enable the reconstruction algorithm to reliably distinguish both the material type and spatial position of the object.

The sensing matrix used for joint detection is constructed as a concatenation of three sub-dictionaries corresponding to different materials, i.e., $\mathbf{H} = [\mathbf{H}_M, \mathbf{H}_N, \mathbf{H}_W]$, where each sub-matrix contains the responses for all spatial positions of metal, nylon, and wood targets, respectively. As a result, each material-position pair is assigned a unique index in the combined dictionary, and the position index ranges from 1 to 303, with each 101-block corresponding to a different material. 

\begin{figure}[t]
    \centering

    \begin{minipage}[t]{0.8\columnwidth}
        \centering
        \rlap{\raisebox{-5pt}{\scriptsize\textbf{(a)}}}
        \includegraphics[width=\linewidth]{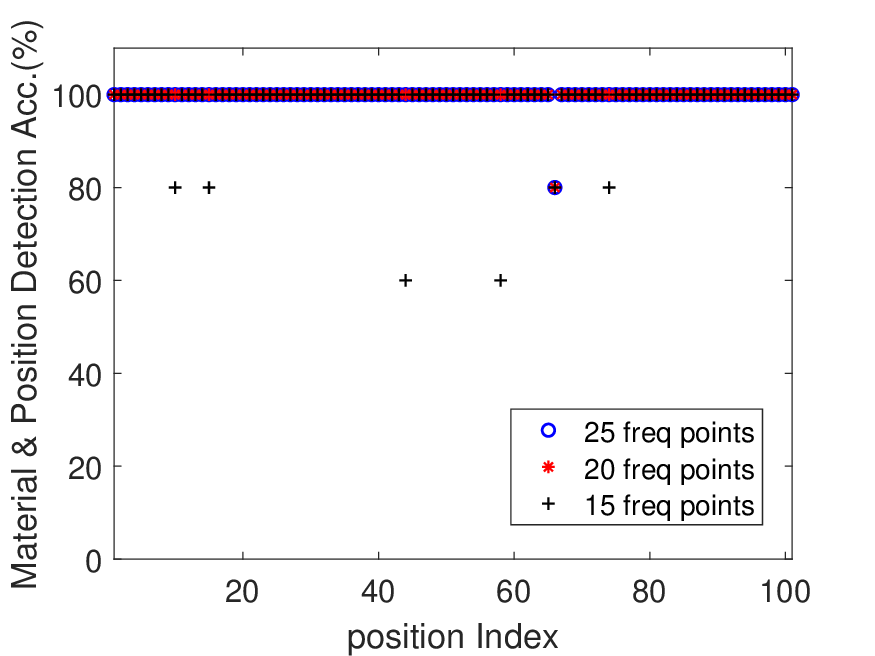}
    \end{minipage}
    \hfill
    \begin{minipage}[t]{0.8\columnwidth}
        \centering
        \rlap{\raisebox{-5pt}{\scriptsize\textbf{(b)}}}
        \includegraphics[width=\linewidth]{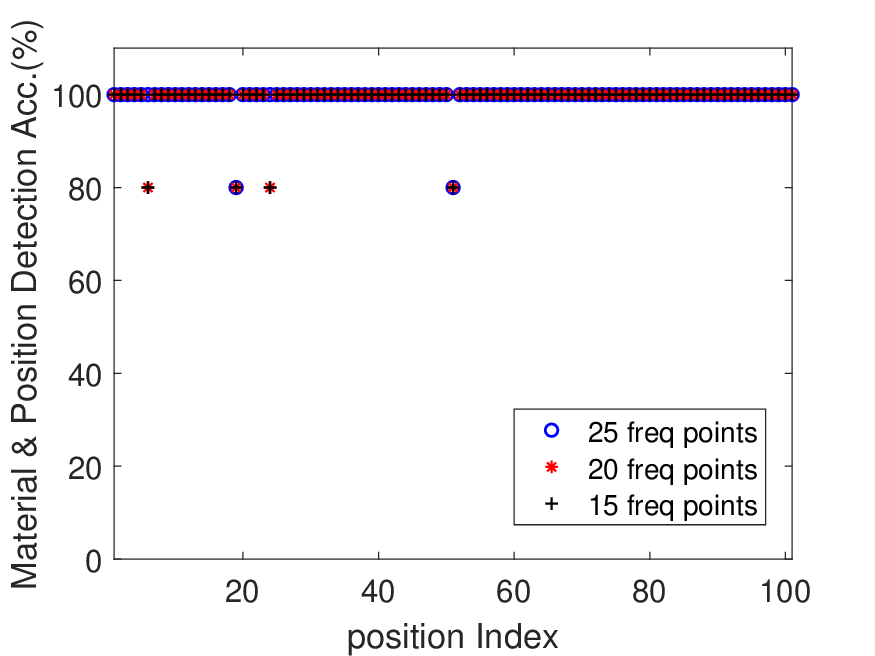}
    \end{minipage}

    \vspace{2pt} 

    \begin{minipage}[t]{0.8\columnwidth}
        \centering
        \rlap{\raisebox{-5pt}{\scriptsize\textbf{(c)}}}
        \includegraphics[width=\linewidth]{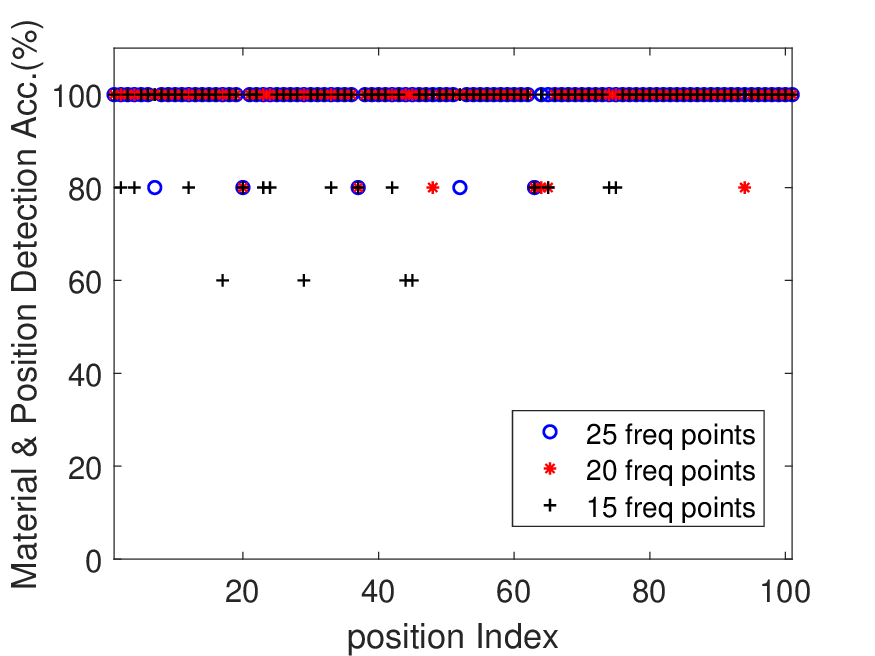}
    \end{minipage}
    \caption{Material and position detection accuracy for targets of (a) metal, (b) nylon, and (c) wood, obtained using 25, 20, and 15 frequency points.}
    \label{fig:fig7}
\end{figure}

First, we examine the possibility of detecting the rod's location and material. The results are shown in Fig. ~\ref{fig:fig7}. Here, a detection is considered correct only when the system correctly identifies both the material type and the target's spatial position. For better visualization, the results are grouped by the true target material to show how joint detection performance varies across object classes. To evaluate the detection accuracy, we examined each of the five measurements available for position–material configuration. Each measurement is treated as an unknown test sample (i.e., $\mathbf{g}$). When all 5 instances are correctly detected, we consider this 100\% accurate. In many practical scenarios, it is desired to obtain the desired accuracy using as narrow a band as possible. As a result, we have reported sensing performance using 25, 20, and 15 frequency points, corresponding to frequency ranges of 10.275--10.635~GHz, 10.275--10.560~GHz, and 10.350--10.560~GHz, respectively. Examining the trends reported in Fig. \ref{fig:fig7}, it is evident that increasing the number of frequency samples improves the overall reliability of the reconstruction for all three materials. In particular, 25 frequency points provide near-perfect performance, whereas reducing the number of frequency points leads to more frequent errors, especially for wooden targets. Since metallic objects are expected to produce stronger perturbations, they are easier to detect and thus require fewer frequency points.

\begin{figure}[t]
    \centering

    \begin{minipage}[t]{0.8\columnwidth}
        \centering
        \rlap{\raisebox{0\height}{\hspace{0mm}\scriptsize\textbf{(a)}}}%
        \includegraphics[width=\linewidth]{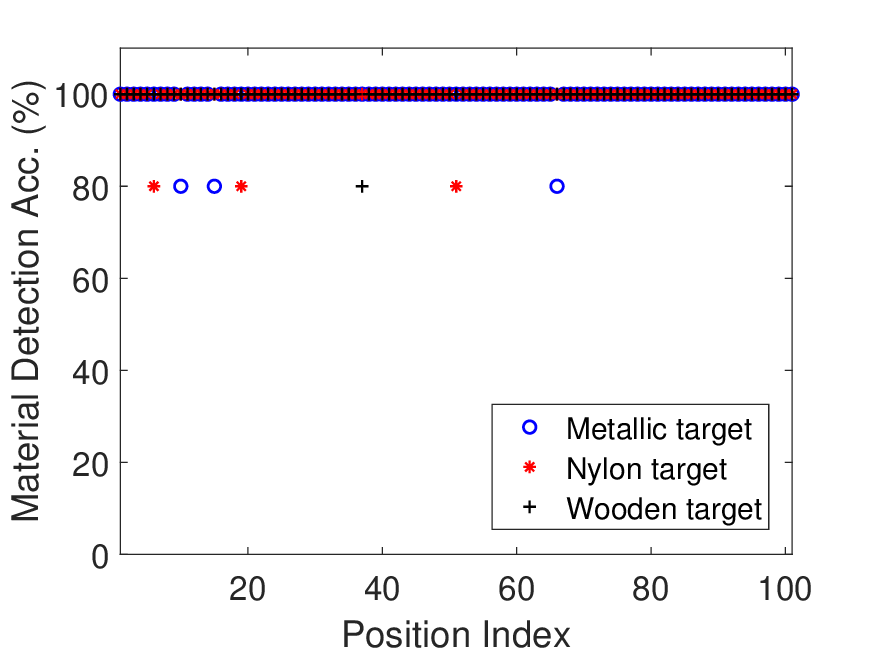}
    \end{minipage}
    \hfill
    \begin{minipage}[t]{0.8\columnwidth}
        \centering
        \rlap{\raisebox{0\height}{\hspace{0mm}\scriptsize\textbf{(b)}}}%
        \includegraphics[width=\linewidth]{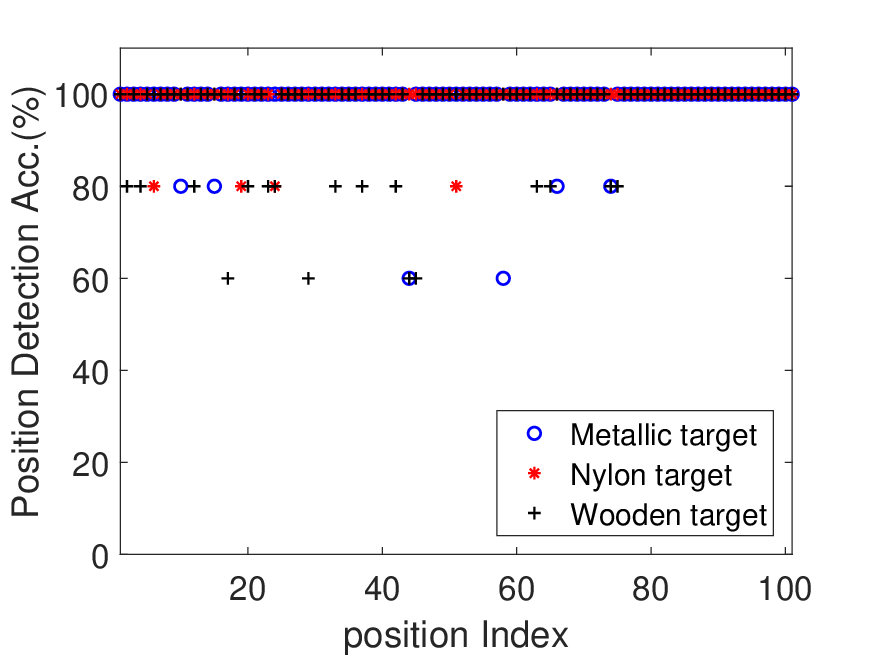}
    \end{minipage}

    \caption{(a) Material-only and (b) position-only detection accuracy obtained using 15 frequency points.}
    \label{fig:fig8}
\end{figure}

\begin{figure}[!t]
\centering
\includegraphics[width=0.8\linewidth]{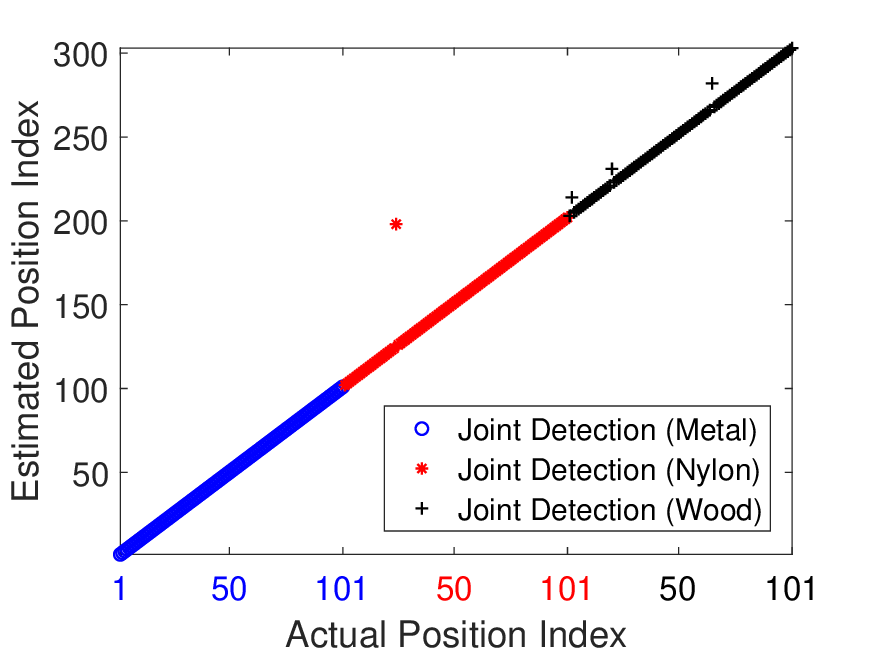}
\caption{Estimated versus actual position indices for joint material and position detection using 15 frequency points.}
\label{fig:figDet}
\end{figure}

To gain additional insight, Fig.~\ref{fig:fig8} separates the joint detection outcome into its two components: material-only accuracy and position-only accuracy, obtained using 15 frequency points. Fig.~\ref{fig:fig8}(a) shows that the material of the target is identified correctly for most test positions, indicating that metal, nylon, and wood produce sufficiently distinct frequency signatures. When compared with Fig.~\ref{fig:fig7}, it is evident that material-only detection achieves much higher accuracy than joint material-and-position detection for the same number of frequency points. Fig.~\ref{fig:fig8}(b), however, shows that position detection is more sensitive to the number of measurements, with more noticeable drops in accuracy at certain locations. These results indicate that, for this sensing framework, material discrimination is generally more robust than precise position localization when the number of frequency samples is limited. As a representative example, Fig.~\ref{fig:figDet} shows the estimated versus actual position indices for one of the five test measurements using 15 frequency points. Correct joint detection of both material and position is indicated by points lying on the diagonal.

\subsection{Detection Performance Without Data Reuse}

In Section III-A, the sensing matrix $\mathbf{H}$ was constructed by averaging all five measurements available for each position--material configuration. While this approach improves robustness to measurement noise, it may introduce a potential bias if the same measurements are used both for constructing $\mathbf{H}$ and for testing. To eliminate any possibility of data leakage, we consider a leave-one-out measurement strategy. For each configuration, the sensing matrix $\mathbf{H}$ is constructed by averaging only four out of the five available measurements, while the remaining measurement is treated as an unseen test sample $\mathbf{g}$. In this way, the test data is fully independent of the sensing matrix.

\begin{figure}[t]
    \centering

    \begin{minipage}[t]{0.8\columnwidth}
        \centering
        \rlap{\raisebox{-10pt}{\scriptsize\textbf{(a)}}}
        \includegraphics[width=\linewidth]{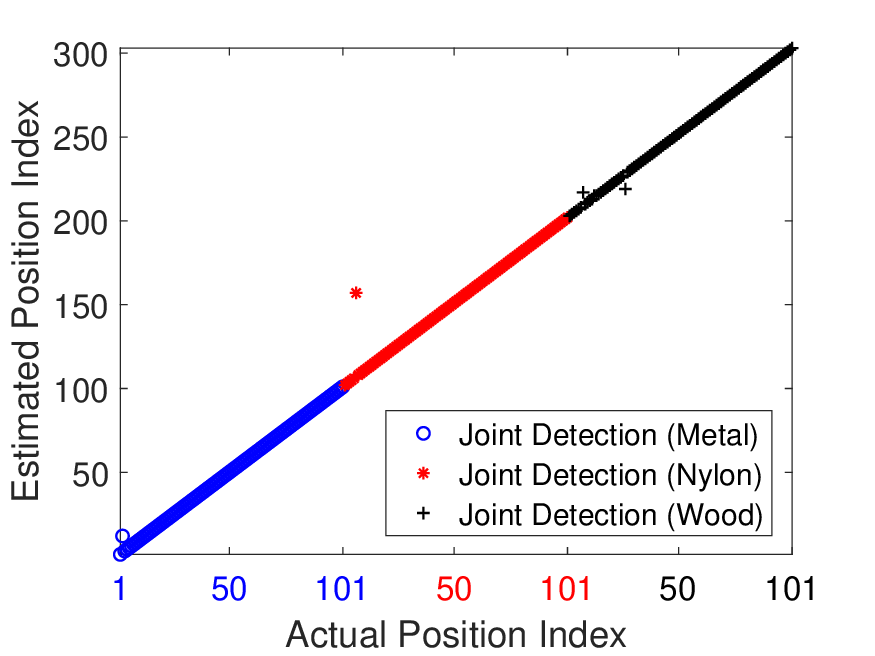}
    \end{minipage}
    \hfill
    \begin{minipage}[t]{0.8\columnwidth}
        \centering
        \rlap{\raisebox{-10pt}{\scriptsize\textbf{(b)}}}
        \includegraphics[width=\linewidth]{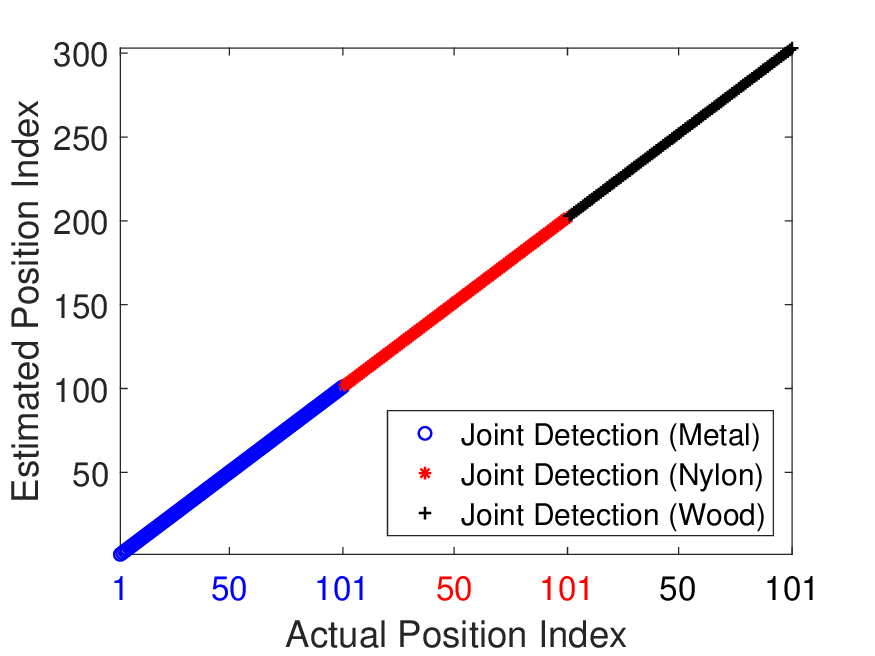}
    \end{minipage}

    \caption{Estimated versus actual position indices for joint material and position detection using the leave-one-out measurement strategy. Results using (a) 25 frequency points. (b) 50 frequency points. The horizontal axis represents the concatenated sensing matrix index, with the tick labels reset from 1 to 101 within each material segment (metal, nylon, and wood).}
    \label{fig:fignew}
\end{figure}

We evaluate the performance of this approach using 25 frequency points corresponding to the same bandwidth as before. The results in Fig.~\ref{fig:fignew}(a) show that the system maintains near-perfect detection accuracy for both material classification and position localization across all test cases. This confirms that the performance reported in Section III-A is not a result of data reuse. Next, we extend the evaluation to a higher number of frequency samples (50 points). In this case, the detection accuracy reaches 100\% for all configurations, demonstrating perfect agreement between the estimated and actual position indices, as illustrated in Fig.~\ref{fig:fignew}(b). The results indicate that increasing the number of frequency points improves the distinctiveness of the sensing signatures and enhances the reliability of the correlation-based detection. These findings validate that the proposed sensing framework generalizes well to unseen measurements and that the constructed sensing matrix remains representative even when fewer samples are used in its formation.

\begin{figure}[t]
    \centering

    \begin{minipage}[t]{0.8\columnwidth}
        \centering
        \rlap{\raisebox{-10pt}{\scriptsize\textbf{(a)}}}
        \includegraphics[width=\linewidth]{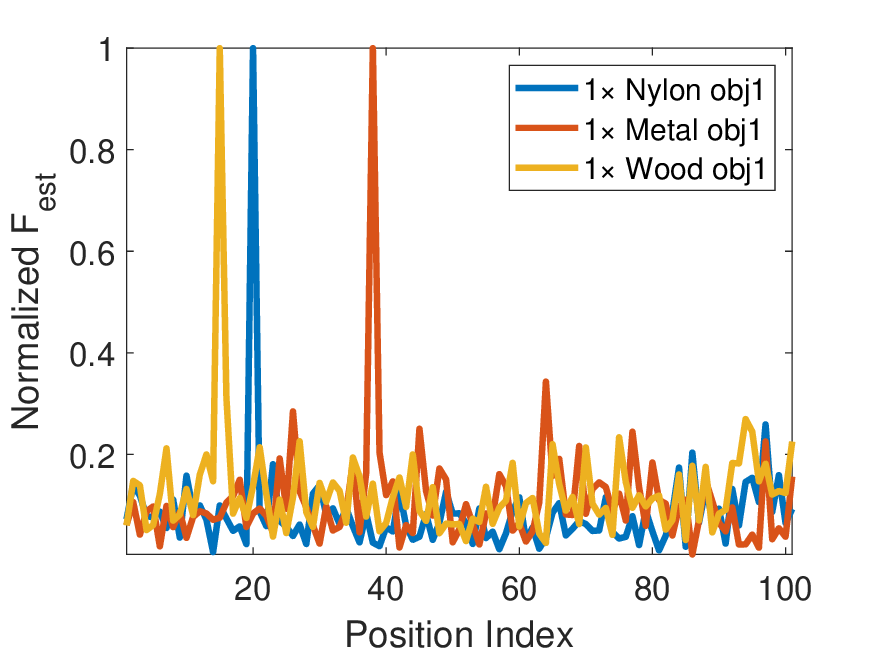}
    \end{minipage}
    \hfill
    \begin{minipage}[t]{0.8\columnwidth}
        \centering
        \rlap{\raisebox{-10pt}{\scriptsize\textbf{(b)}}}
        \includegraphics[width=\linewidth]{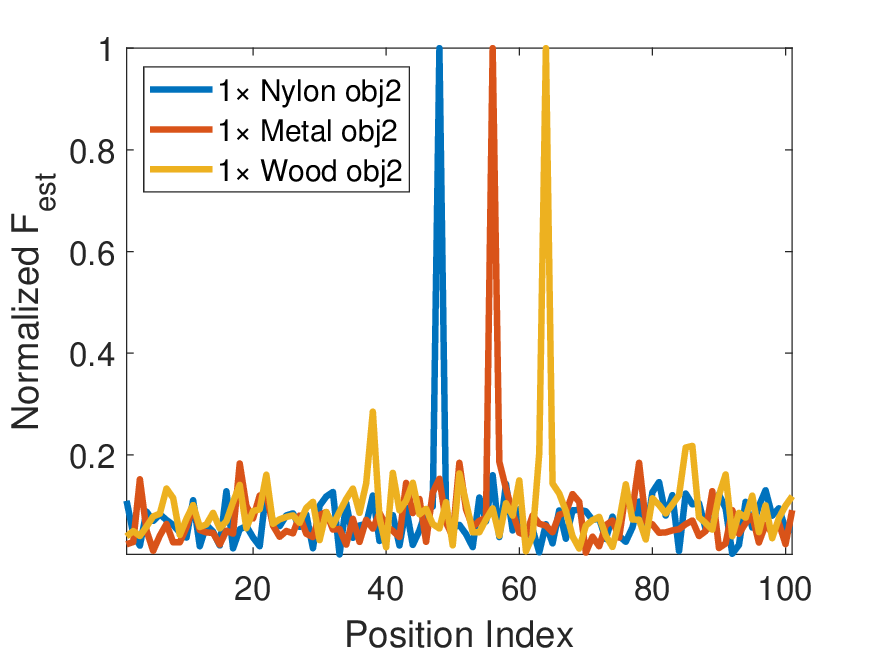}
    \end{minipage}

    \vspace{2pt} 

    \begin{minipage}[t]{0.8\columnwidth}
        \centering
        \rlap{\raisebox{-10pt}{\scriptsize\textbf{(c)}}}
        \includegraphics[width=\linewidth]{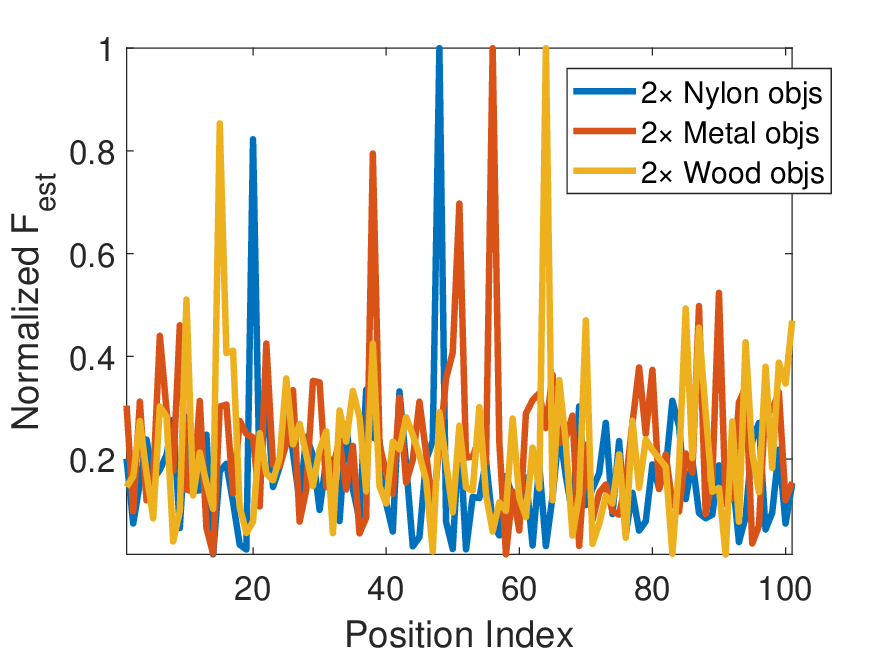}
    \end{minipage}

    \caption{Normalized $|\mathbf{F}_{\text{est}}|$ for (a) and (b) Single object placed inside the pipe at different locations and (c) two objects placed at locations in (a) and (b).}
    \label{fig:fig9}
\end{figure}

\subsection{Two-Object Detection}

To further evaluate the sensing capability, two rods are simultaneously placed at different grid locations. In this case, we aim to detect the two rods, assuming that the measured signal is a linear superposition of the corresponding columns of the sensing matrix. Because of multiple scattering, this assumption is only a reliable approximation for weak scatterers \cite{delhougne2018precise}. Furthermore, detecting two objects is a more complicated inverse problem. Because of these challenges, localizing two objects requires more frequency samples (i.e. larger bandwidth). For this reason, 250 frequency points (8.250 to 11.985 GHz) are used here to provide sufficient measurement diversity and improve the stability of the reconstruction. The inverse problem is then solved using the CGS algorithm, which is well-suited to non-square and potentially ill-conditioned sensing matrices. As a result, the reconstruction process produces two distinct peaks in the estimated vector $|\mathbf{F}_{\text{est}}|$, each indicating the presence and positions of one of the objects, as shown in Fig.~\ref{fig:fig9}(c). For comparison, Fig.~\ref{fig:fig9}(a) and (b) show the estimated vector for the single-object cases, where only one dominant peak is observed. These results demonstrate that the proposed conformal metasurface sensing system may also be used to detect multiple (small/weak) scatterers within the sensing region.

\section{Conclusion}
This paper presented a compact microwave sensing framework based on conformal frequency-diverse metasurface antennas for joint material identification and spatial localization within a cylindrical domain. By leveraging frequency multiplexing, the proposed approach eliminates the need for antenna arrays, switching networks, and mechanical scanning, reducing system complexity while maintaining effective sensing capability. A sensing matrix constructed from background-subtracted transmission response measurements enables reliable reconstruction of both material type and object location using computational processing. The experimental results demonstrate high detection accuracy across different materials and show improved performance with increased frequency diversity. We also report proof-of-concept results for detecting two targets within the sensing region. These results highlight the potential of conformal metasurface-based sensing as a practical and scalable solution for non-destructive evaluation and smart sensing applications. In the future, the metasurfaces will be redesigned to exhibit improved reflection coefficient, and radiate into dense material (e.g. concrete) to detect or localize changes for NDE of infrastructure, human gesture recognition \cite{kashyap2023leaky}, and biomarker sensing.

\section*{Acknowledgment}

This material is based upon work supported by the National Science Foundation under Grant No. ECCS-2333023.

\bibliographystyle{IEEEtran}
\bibliography{reference}

\vfill

\end{document}